\documentclass{llncs}

\usepackage[hyphens]{url}
\usepackage{cite}
\usepackage{graphicx}
\usepackage{framed}
\usepackage{breqn}
\usepackage{amsfonts}
\usepackage{listings}
\usepackage{color}

\definecolor{mygreen}{rgb}{0,0.2,0}
\definecolor{mygray}{rgb}{0.95,0.95,0.95}
\definecolor{mymauve}{rgb}{0.58,0,0.82}

\definecolor{lbcolor}{rgb}{0.95,0.95,0.95}

\lstdefinelanguage{cindyjs}{
  keywords={draw, complex, sqrt, join, gauss, meet, floor, colorplot, exp, drawcircle, drawtext},
  ndkeywords={>>},
  sensitive=true
}

\begin{document}

\title{On Euler's inequality and \\ automated reasoning with dynamic geometry}
\author{Zolt\'an Kov\'acs\orcidID{0000-0003-2512-5793}\inst{1} \and R\'obert Vajda\inst{2} \and Aaron Montag\inst{3}}
\institute{
The Private University College of Education of the Diocese of Linz\\
Salesianumweg 3, A-4020 Linz, Austria\\
\email{zoltan@geogebra.org} \and
Bolyai Institute, University of Szeged\\
Aradi v\'ertan\'uk tere 1, H-6720 Szeged, Hungary\\
\email{vajdar@math.u-szeged.hu} \and
Technical University of Munich, Germany\\
Arcisstra\ss e 21, D-80333 M\"unchen\\
\email{montag@ma.tum.de}
}

\maketitle              

\begin{abstract}
Euler's inequality $R\geq 2r$ can be investigated
in a novel way by
using implicit loci in GeoGebra. Some unavoidable side effects of the implicit locus computation
introduce unexpected algebraic curves. By using a mixture of symbolic and numerical
methods a possible approach is sketched up to investigate the situation.
By exploiting fast GPU computations, a web application written in CindyJS helps
in understanding the situation even better.

\end{abstract}
\section{GeoGebra: a symbolic tool for obtaining generalizations of geometric statements}
\textit{GeoGebra} \cite{gg} is a well known dynamic geometry software package with millions
of users worldwide. Its main purpose is to visualize geometric invariants.
Recently GeoGebra has been supporting investigation of geometric constructions
also symbolically by harnessing the strength of the embedded computer algebra system (CAS) \textit{Giac}
\cite{GiacGG-RICAM2013}. One direct use of the embedded CAS is automated reasoning
\cite{gg-art-doc-gh}. In this paper we use in particular the implicit locus
derivation feature \cite{ART-ISSAC2016} in GeoGebra, by using the command
\texttt{LocusEquation} with two inputs: a Boolean expression and the sought mover point.
For example, given an arbitrary triangle $ABC$ with sides $a$, $b$ and $c$,
entering \texttt{LocusEquation[$a$==$b$,$C$]} results in the perpendicular bisector $d$ of $AB$,
that is, if $C$ is chosen to be an element of $d$, then the condition $a=b$ is satisfied.

Obtaining implicit loci is a recent method in GeoGebra to get interesting facts on classic theorems.
These facts are closely related to algebraic curves which usually describe generalization of
the classic results. Sometimes it is computationally difficult to obtain the curves
quickly enough, but some new improvements in Giac's elimination algorithm opened
the road to effectively investigate a large number of geometric constructions \cite{mcs,ACA2017}
including \textit{Holfeld's 35th problem} \cite{mcs,ACA2015}, a generalization of the
\textit{Steiner-Lehmus theorem} \cite{mcs,SteinerLehmus} or \textit{the right triangle altitude theorem}
\cite{ART-ISSAC2016}.

We need to admit that the possibility to generalize well known theorems is a consequence
of using \textit{unordered geometry} \cite[p.~97]{chou} in the applied tools and theories. In unordered geometry
one cannot designate only one intersection point of a line and a conic (or two conics),
so \textit{both} will be considered at the same time. (See \cite[p.~59]{chou} for an example
on irreducible problems and undistinguishable cases.) Therefore we obtain a larger
set of points for the resulting algebraic curve as expected. The obtained set may be inconvenient
in some cases, but can be fruitful to get some interesting generalizations.

Finally we demonstrate a new approach in using the GPU of the user's computer by
utilizing \textit{CindyJS} \cite{cindyjs} that colors the points of the plane according to a predefined relationship.
In this way we can have a numerical solution very quickly, however some \textit{CindyScript} programming
will be required.

\section{Euler's inequality}

We recall that in Euclidean planar geometry Euler's inequality states that 
$R\geq 2r$ where $R$ and $r$ denote the radius of the circumscribed circle and the inscribed circle of a triangle,
respectively.

\begin{figure}
\begin{center}
\includegraphics[scale=0.35]{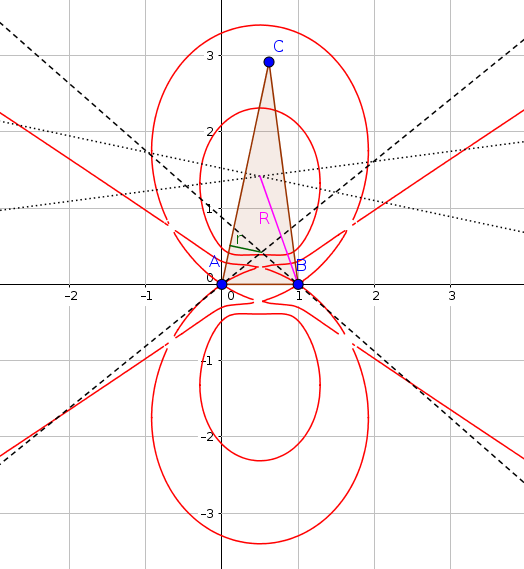}
\caption{An implicit curve (in red) as the output of GeoGebra command \texttt{LocusEquation[$R$==$3r$,$C$]}.
Here points $A$ and $B$ are fixed in the plane and $C$ is a free point. (In other words: Triangle $ABC$ has
fixed vertices $A$ and $B$.)
The computation of the sought set of mover point $C$ is a moderately
difficult problem: an Intel(R) Core(TM) i7 CPU 860 @ 2.80GHz computer requires 3.7 seconds
for the whole computation.}
\label{fig:Euler3}
\end{center}
\end{figure}

Since GeoGebra's \textit{Automated Reasoning Tools} \cite{gg-art-doc-gh} use Gr\"obner bases in the background,
inequalities cannot really be investigated by them automatically.\footnote{
Here we refer to \cite[p.~227]{recio-dalzotto} which suggests using a different approach, based purely on equations by
investigating the distance of the centers of the circumscribed and inscribed circles.}
Certain
experiments can however be started by fixing the ratio of the studied quantities,
here $R$ and $r$.
For example, one can start with some concrete experience by comparing $R$ and say $3r$ (see
Fig.~\ref{fig:Euler3}), and then simply change the constant $3$ to some different value.
As output, the red curve in the figure gives a necessary geometric condition where to put $C$ in order to have $R=3r$.

The result seems complicated for the first look. By doing some more experiments, it turns out that the
two inner oval parts of the curve show relevant information on the concrete question, but the other parts show something different.
That is, by setting $C$ to an arbitrary point of the inner oval parts, the equality $R=3r$ will occur.
For the other parts we will see later in Sec.~\ref{subsec:octic} that the radii $r_a$, $r_b$ and $r_c$ of the excircles will
take the role of $r$ over.

\begin{figure}
\begin{center}
\includegraphics[scale=0.35]{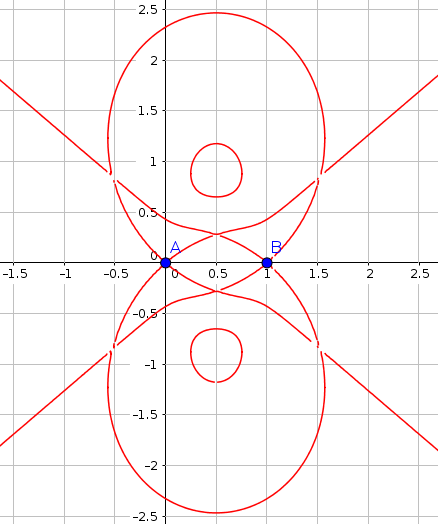}
\includegraphics[scale=0.35]{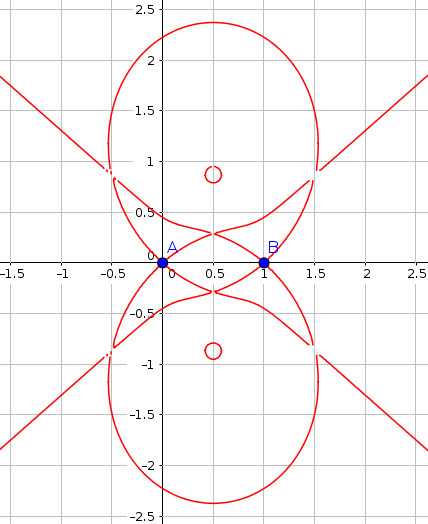}
\caption{Result of \texttt{LocusEquation[$R$==$2.1r$,$C$]} and \texttt{LocusEquation[$R$==$2.01r$,$C$]}.
To properly plot the latter a suitable zoom factor may be required due to possible inaccuracies
in the plotting routine in GeoGebra.}
\label{fig:euler-2plus}
\end{center}
\end{figure}

\begin{figure}
\begin{center}
\includegraphics[scale=0.35]{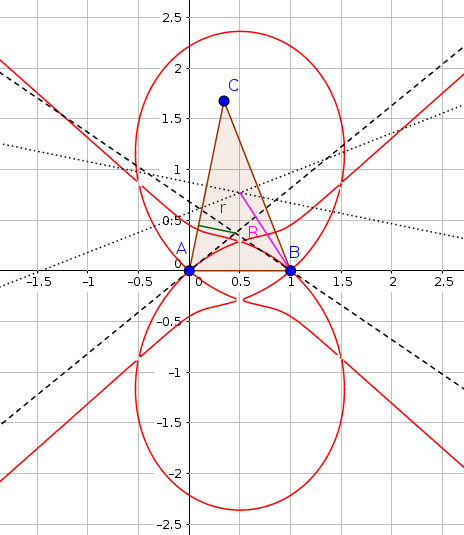}
\caption{The inner oval parts disappear when plotting the case $R=2r$.}
\label{fig:Euler2}
\end{center}
\end{figure}

After doing further experiments by changing the constant $3$ to lower values, when getting close to $2$ the
inner oval parts seem to disappear even more and more (Fig.~\ref{fig:euler-2plus}),
and finally for the experiment $R=2r$ the inner oval parts
are not visible any longer (Fig.~\ref{fig:Euler2}).

The first confusing result is why the points $\left(\frac{1}{2},\pm\frac{\sqrt{3}}{2}\right)$
are not plotted in this graph---we recall that the
equality holds if and only if the triangle is equilateral.
Unfortunately, the plotting routine in GeoGebra does not
show this isolated point. In fact, other systems (including \textit{Wolfram$\mid$Alpha} and \textit{Desmos})
are also unable to automatically plot even the easiest examples of a very similar
situation, namely that a curve has an \textit{acnode}. Such a basic example is the curve $x^3-x^2-y^2=0$
for which the point $(0,0)$ is not shown in the graph, but is clearly an isolated point of the curve
\cite{acnode}.

The result of the command \texttt{LocusEquation[$R$==$2r$,$C$]} is
\begin{dmath*}
x^{18} - 225 \; y^{18} - 1481 \; x^{2} \; y^{16} - 4004 \; x^{4} \; y^{14} - 5460 \; x^{6} \; y^{12} - 3262 \; x^{8} \; y^{10} + 770 \; x^{10} \; y^{8} + 2604 \; x^{12} \; y^{6} + 1756 \; x^{14} \; y^{4} + 535 \; x^{16} \; y^{2} - 504 \; x^{17} + 1256 \; x \; y^{16} + 6752 \; x^{3} \; y^{14} + 13632 \; x^{5} \; y^{12} + 10336 \; x^{7} \; y^{10} - 4400 \; x^{9} \; y^{8} - 14304 \; x^{11} \; y^{6} - 11008 \; x^{13} \; y^{4} - 3808 \; x^{15} \; y^{2} + 1764 \; x^{16} + 1276 \; y^{16} + 1416 \; x^{2} \; y^{14} - 6936 \; x^{4} \; y^{12} - 11544 \; x^{6} \; y^{10} + 8608 \; x^{8} \; y^{8} + 33048 \; x^{10} \; y^{6} + 30104 \; x^{12} \; y^{4} + 11896 \; x^{14} \; y^{2} - 3528 \; x^{15} - 2888 \; x \; y^{14} - 2296 \; x^{3} \; y^{12} + 6296 \; x^{5} \; y^{10} - 7000 \; x^{7} \; y^{8} - 41944 \; x^{9} \; y^{6} - 47080 \; x^{11} \; y^{4} - 21368 \; x^{13} \; y^{2} + 4410 \; x^{14} - 1094 \; y^{14} + 2854 \; x^{2} \; y^{12} - 1262 \; x^{4} \; y^{10} - 370 \; x^{6} \; y^{8} + 31246 \; x^{8} \; y^{6} + 46226 \; x^{10} \; y^{4} + 24230 \; x^{12} \; y^{2} - 3528 \; x^{13} - 2888 \; x \; y^{12} + 592 \; x^{3} \; y^{10} + 5704 \; x^{5} \; y^{8} - 12704 \; x^{7} \; y^{6} - 29240 \; x^{9} \; y^{4} - 17840 \; x^{11} \; y^{2} + 1764 \; x^{12} + 1276 \; y^{12} - 1136 \; x^{2} \; y^{10} - 5940 \; x^{4} \; y^{8} + 1472 \; x^{6} \; y^{6} + 11604 \; x^{8} \; y^{4} + 8368 \; x^{10} \; y^{2} - 504 \; x^{11} + 1256 \; x \; y^{10} + 2984 \; x^{3} \; y^{8} + 912 \; x^{5} \; y^{6} - 2608 \; x^{7} \; y^{4} - 2296 \; x^{9} \; y^{2} + 63 \; x^{10} - 225 \; y^{10} - 581 \; x^{2} \; y^{8} - 330 \; x^{4} \; y^{6} + 246 \; x^{6} \; y^{4} + 283 \; x^{8} \; y^{2}=0.
\end{dmath*}
By using GeoGebra's \texttt{Substitute[$I$,\{$x$=1/2,$y$=sqrt(3)/2\}]} command (here $I$ denotes
the obtained implicit curve object) we get $0=0$ which shows
that the expected point is indeed an element of the curve.
The same result can be seen for the point $\left(\frac{1}{2},-\frac{\sqrt{3}}{2}\right)$.

The obtained polynomial can be factored by using GeoGebra's \texttt{Factor[LeftSide[$I$]-RightSide[$I$]} command.
The factorization is
\begin{dmath*}
\left(x^{2} + y^{2} \right) \; \cdot \;  \left(7 \; x^{8} - 28 \; x^{7} + 12 \; x^{6} \; y^{2} + 42 \; x^{6} - 36 \; x^{5} \; y^{2} - 28 \; x^{5} - 6 \; x^{4} \; y^{4} + 34 \; x^{4} \; y^{2} + 7 \; x^{4} + 12 \; x^{3} \; y^{4} - 8 \; x^{3} \; y^{2} - 20 \; x^{2} \; y^{6} - 26 \; x^{2} \; y^{4} - 2 \; x^{2} \; y^{2} + 20 \; x \; y^{6} + 20 \; x \; y^{4} - 9 \; y^{8} + 46 \; y^{6} - 9 \; y^{4} \right) \; \cdot \; \left(9 \; x^{8} - 36 \; x^{7} + 52 \; x^{6} \; y^{2} + 54 \; x^{6} - 156 \; x^{5} \; y^{2} - 36 \; x^{5} + 102 \; x^{4} \; y^{4} + 190 \; x^{4} \; y^{2} + 9 \; x^{4} - 204 \; x^{3} \; y^{4} - 120 \; x^{3} \; y^{2} + 84 \; x^{2} \; y^{6} + 186 \; x^{2} \; y^{4} + 34 \; x^{2} \; y^{2} - 84 \; x \; y^{6} - 84 \; x \; y^{4} + 25 \; y^{8} - 14 \; y^{6} + 25 \; y^{4} \right).
\end{dmath*}
Here the first factor $p_1=x^2+y^2$ clearly corresponds to the point $A$.
The second factor $p_2=7x^8-\ldots$
shows all real points of the curve $I$
(without the points $\left(\frac{1}{2},\pm\frac{\sqrt{3}}{2}\right)$), and the third factor $p_3=9x^8-\ldots$
has seemingly no real points, but after
computing its acnodes by
solving the inequality system $p_3=0$, $(p_3)'_x=0$, $(p_3)'_y=0$, $H(p_3)>0$, where
$H$ denotes the Hessian matrix, we may explore symbolically that the polynomial indeed describes the
two expected isolated real points as well.

\begin{figure}
\begin{center}
\includegraphics[width=0.7\textwidth]{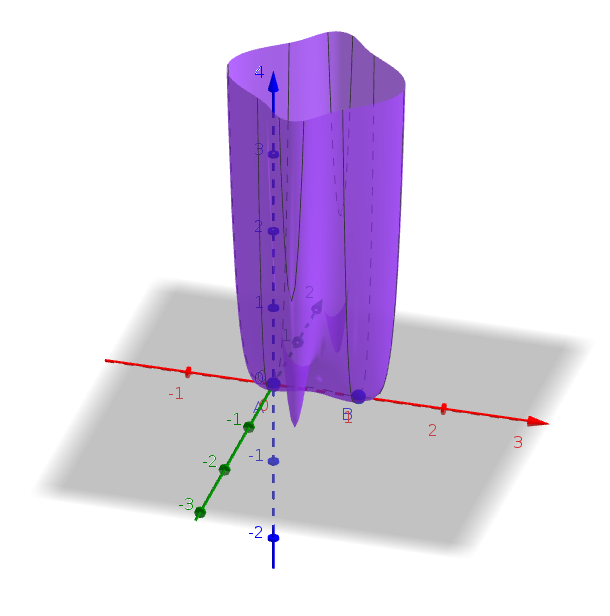}
\caption{A 3D plot of $p_3$ in GeoGebra. Here we used the command
\texttt{$f(x,y):=$Element[Factors[LeftSide[$I$]-RightSide[$I$]],3,1]} and
opened the Graphics 3D View.}
\label{fig:Euler2-f3-3d}
\end{center}
\end{figure}

\begin{figure}
\begin{center}
\includegraphics[width=0.7\textwidth]{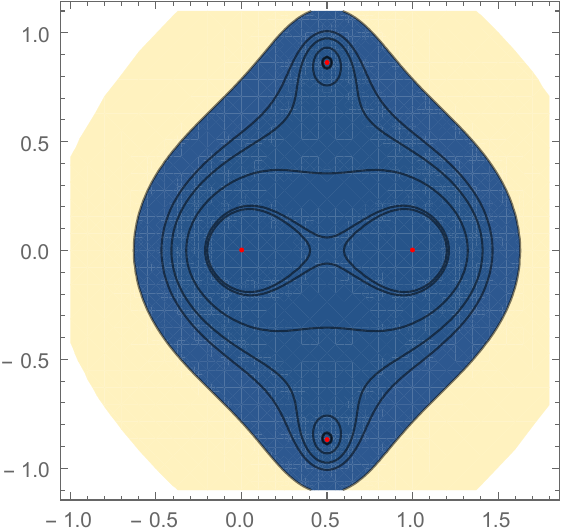}
\caption{A contour plot of $p_3$ in \textit{Mathematica}.}
\label{fig:conteuler}
\end{center}
\end{figure}

\begin{figure}
\begin{center}
\includegraphics[width=0.7\textwidth]{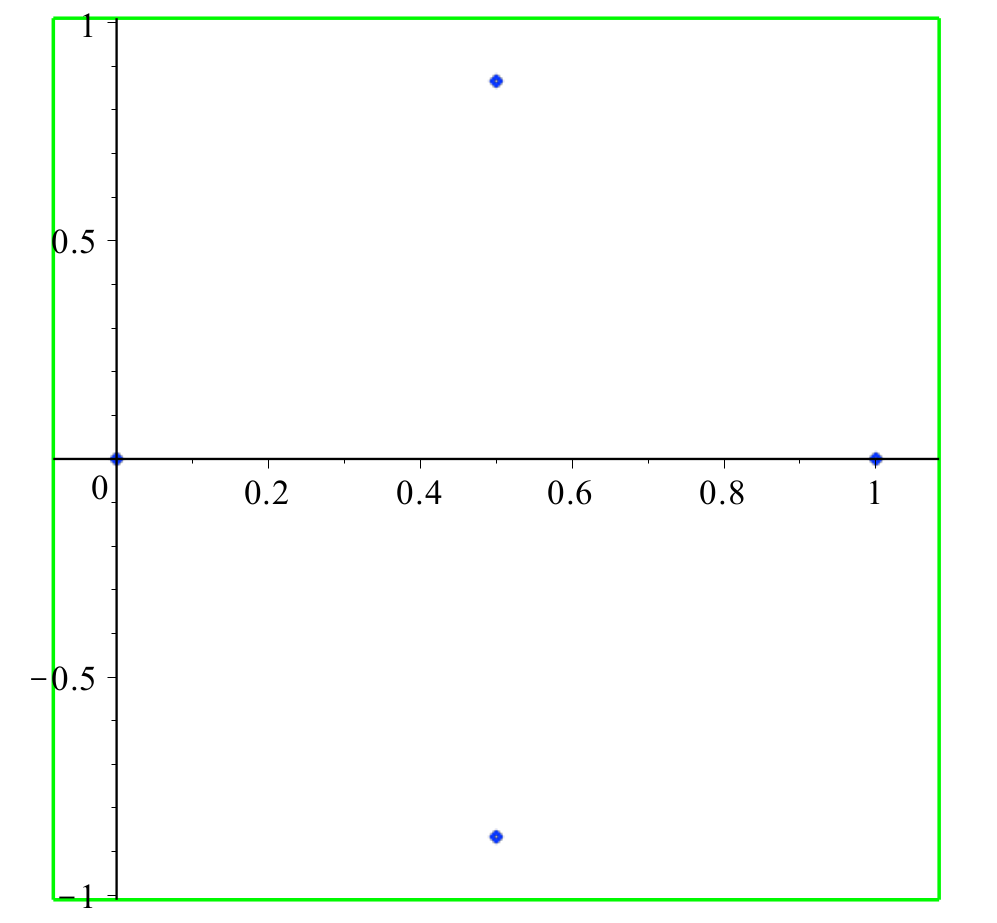}
\caption{Plotting $p_3$ with Maple's \texttt{algcurves} package and its \texttt{plot\_real\_curve} command.}
\label{fig:maple-real}
\end{center}
\end{figure}

This approach with the Hessian cannot be achieved in GeoGebra.
Instead, a numerical way can be tried to visualize the function $f(x,y)=p_3$ in 3 dimensions
(Fig.~\ref{fig:Euler2-f3-3d}) to find the real roots, namely $\left(\frac{1}{2},\pm\frac{\sqrt{3}}{2}\right)$,
$(0,0)$ and $(1,0)$. Also in some other computer algebra systems a contour plot may help (see
Fig.~\ref{fig:conteuler}), or to use extra packages which have more sophistical methods to plot real curves
(as seen in Fig.~\ref{fig:maple-real}).

Finally we remark that by using
\textit{Maple}'s \texttt{evala(AFactor($\ldots$))} command we can verify that $p_2$ and $p_3$
are irreducible over $\mathbb{C}$. This can also be achieved by using Singular's absolute factorization library
(\texttt{absfact\_lib}).

\subsection{Summary of the difficulties}

The above shows some difficulties in our case. First of all, by using Gr\"obner bases there seems no
completely automatic way to obtain Euler's inequality---however, the paper \cite{recio-dalzotto} sketches up a
possible method (without full explanation in general). In our approach one needs to start some experiments by choosing
the ratio between $R$ and $r$ randomly. In our opinion, this problem can be automatically resolved by using real geometry
and quantifier elimination \cite{CollinsHong} not only in our case, but in general (see \cite{Robu} for some details).

The second problem is that the plotted graph can be inaccurate: the
equilateral case for $R=2r$ cannot be read off by the user in GeoGebra. It would be expected that the output curve should \textit{contain}
the set of points where the equality holds---this does not seem to be the case here because of the
failure of the plotting algorithm. The case of failure even for some easy cubic examples show
that this problem cannot be easily worked around without using extra software packages.

For similar reasons the factorization does not directly help finding the equilateral case, either. Only
a 3D plot---actually a numerical approach---gives some hints where to look for the equality.

\subsection{Why the octic $p_2$?}
\label{subsec:octic}

Similarly to the Steiner-Lehmur generalization in \cite{SteinerLehmus} here we silently introduced
three other circles as extensions of the incircle. They are the excircles---in unordered
geometry one cannot distinguish between internal and external angle bisectors.
(See also \cite[p.~60]{chou} for a discussion on this.)

\begin{figure}
\begin{center}
\includegraphics[width=0.7\textwidth]{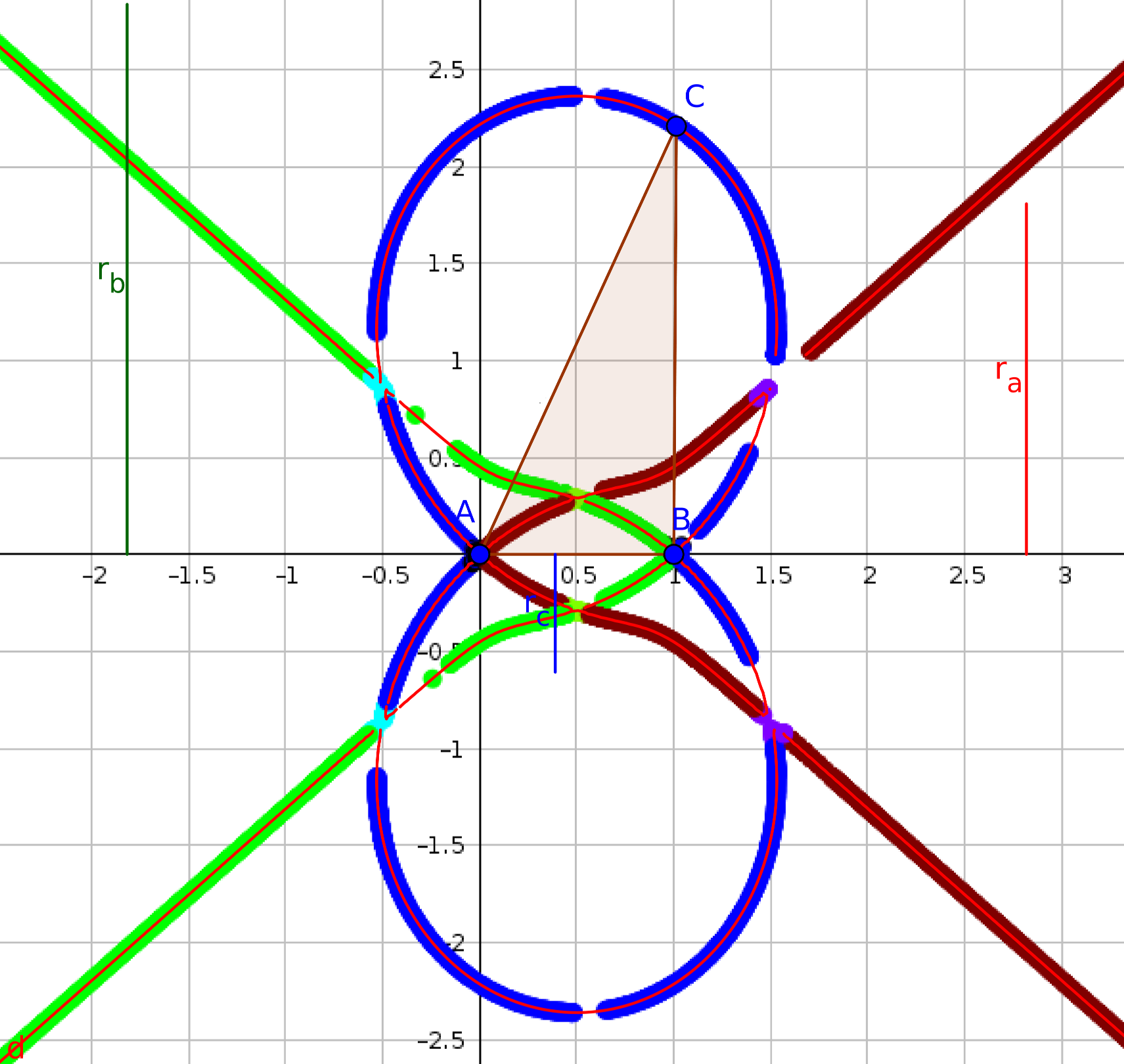}
\caption{Various parts of the octic $p_2$ show the sought moving points $C$ for statements
$R=2r_a$ (red), $R=2r_b$ (green) and $R=2r_c$ (blue). The figure was produced with
GeoGebra by attaching a new point (denoted again by $C$) to the octic $p_2$,
and then constructing $r_a$, $r_b$, $r_c$ and $R$
by using the attached point as ``$C$'', and computing which excircle would be connected
with the appropriately chosen attached point---then the color of $C$ was dynamically
set by using the RGB scheme (see \cite{extraverification} for a similar concept).
Finally \textit{tracing} and \textit{animation} were switched on for $C$ to cover
all possible screen points of the octic.}
\label{fig:euler-ineq-parts2}
\end{center}
\end{figure}

After some experimenting it can be concluded that different sections of the octic $p_2$ describe different
circles among the three excircles (Fig.~\ref{fig:euler-ineq-parts2}).

\subsection{The inequality does not hold for excircles}

\begin{figure}
\begin{center}
\includegraphics[scale=0.35]{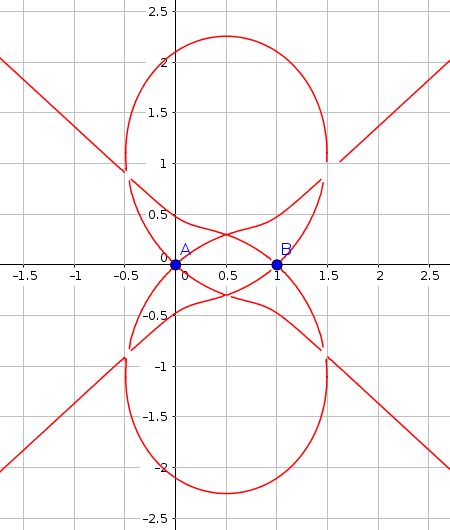}
\caption{Result of \texttt{LocusEquation[$1.9r$==$R$,$C$]}.}
\label{fig:euler-2minus}
\end{center}
\end{figure}

Continuing the process that changing the constant $3$ to lower values, including less
numbers than $2$, we learn that the inner oval parts of the curve will not be visible
any longer. This is the case e.~g.~for $1.9$: there are no visible inner oval parts (and they do not
exist, either, because of Euler's inequality), but the other parts still do (Fig.~\ref{fig:euler-2minus}). This
supports the idea that the inequality with respect to $r$ cannot be transferred to
$r_a$, $r_b$ or $r_c$. That is, we concluded that Euler's inequality always fails
on excircles.

\section{Another numerical approach: CindyJS}

CindyJS \cite{cindyjs} is a JavaScript library and built on top of the
NodeJS infrastructure. It is designed to be able to load dynamic geometry
files produced by Cinderella \cite{cinderella} in the long term.  The
plugin CindyGL utilizes the WebGL subsystem of the user's web browser and
gives direct access to the GPU to exploit fast computations.
As a result
various mathematical formulas can be investigated in a real-time way to
get immediate feedback on the changing input when the user drags input points,
for example. 
CindyJS has been developed by a team of computer graphics and dynamic
geometry experts under the leadership of the Technical University of
Munich, Germany.

By using a web browser and its access to the GPU
a user can do fast visualization of the above properties. The following program
code, shown and explained in shorter parts, is capable to use the vertices of a triangle $ABC$ as input, and then compute its
radius $r$ and $R$ of the incircle and the circumcircle, respectively. Finally
a contour plot is drawn to classify points of the plain with respect to the ratio $R/r$
in case the given point of the plane is chosen as $C$.

\begin{lstlisting}
angularbisectors(P, Q, R) := (
  p = complex(P);
  q = complex(Q);
  r = complex(R);
  w = sqrt((r-p)*(q-p));
  join(P, gauss(p+w));
);
\end{lstlisting}
The function \texttt{angularbisectors} returns the internal bisector
of $\angle QPR$. The command \texttt{gauss} converts a complex number to a
point and the command \texttt{join} connects its input points to represent the result as a line---this
last command will be returned by the function.

To compute the distance from point $P$ to line $l$ we use the function
\begin{lstlisting}
dist(P, l) := |P_1*l_1+P_2*l_2+P_3*l_3|/(|(l_1,l_2)|*|P_3|);
\end{lstlisting}

We also need to create the perpendicular bisector of the input points $P$ and $Q$:
\begin{lstlisting}
perpbisector(P, Q):= (
  p = complex(P);
  q = complex(Q);
  mid = (p+q)/2;
  join(gauss(mid), gauss(mid+i*(q-p)))
);
\end{lstlisting}

Now we are ready to map a color to each point of the plane. We assume that for all $C$
points in the plane the function $f$ is to be used:
\begin{lstlisting}
f(C) := (
  b1 = angularbisectors(A, B, C);
  b2 = angularbisectors(B, C, A);
  incenter = meet(b1, b2);
  r = dist(incenter, join(A, B));
  p1 = perpbisector(A, B);
  p2 = perpbisector(B, C);
  circumcenter = meet(p1, p2);
  R = |circumcenter.xy - A.xy|;
  levels = 30;
  floor(0.01/(R/r-2)*levels)/levels;
);
\end{lstlisting}
Here \texttt{meet} computes the intersection point of the input lines.
After computing the values of $r$ and $R$ the final command designates the
returned color for each input point $C$ by computing a numerical contour plot.

\begin{figure}
\begin{center}
\includegraphics[width=0.6\textwidth]{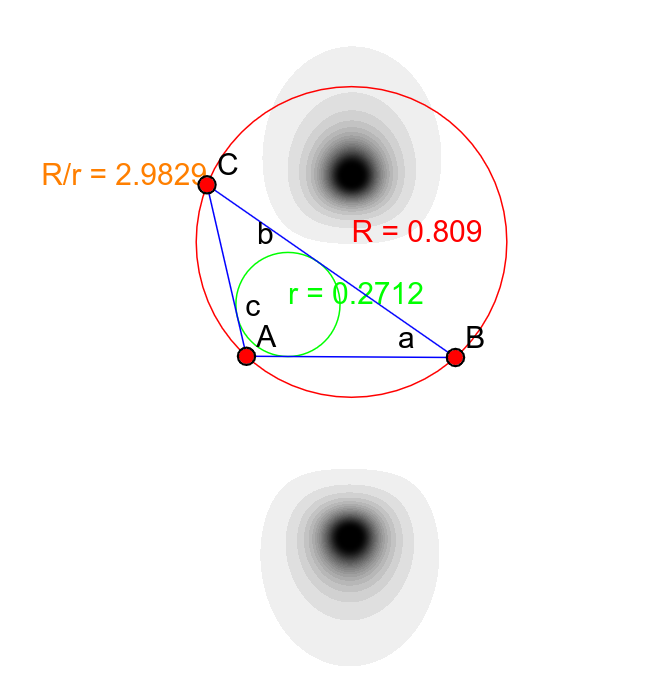}
\caption{Output of CindyJS.}
\label{fig:euler-cindyjs}
\end{center}
\end{figure}

Finally we use the \texttt{colorplot} function with the running variable \texttt{\#}
that stands for all possible values of $C$ in the plane: 
\begin{lstlisting}
colorplot(exp(-2*|f(#)|));
\end{lstlisting}

To show only one particular value of $C$, that is, the current value that is based
on the position of the point $C$ in the triangle being shown, we simply use
\begin{lstlisting}
f(C);
\end{lstlisting}
and thereafter we will be able to draw the incircle and the circumcircle and print their
radii on the screen:
\begin{lstlisting}
drawcircle(incenter, r, color->[0,1,0]);
drawtext(incenter, "r = " + r, color->[0,1,0]);
drawcircle(circumcenter, R, color->[1,0,0]);
drawtext(circumcenter, "R = " + R, color->[1,0,0]);
drawtext(C, "R/r = " + R/r, color->[1,.5,0], align->"right");
\end{lstlisting}

The output of the code can be seen in Fig.~\ref{fig:euler-cindyjs}. It is immediately clear
that the ratio $R/r$ has an extremum in the ``black area'' of the figure. The black layer
corresponds to ratios between $2$ and $2+\varepsilon$, the next level (that is a little bit lighter)
corresponds to ratios between $2+\varepsilon$ and $2+2\varepsilon$, and so on. The value of $\varepsilon$
can be fine tuned at the end of the declaration of \texttt{f(C)} above.

We highlight that the output can be better observed when points $A$ and $B$ are dragged. In this way we
can ``zoom in'' the figure and have a more detailed understanding on the geometry of Euler's inequality.

The full source
code of this example can be found at \url{github.com/CindyJS/CindyJS/blob/master/examples/cindygl/63_eulerinequality.html}.
On further details on programming CindyJS we refer to \cite{cindyjs2}.

\section{Conclusion}

We used a novel method to obtain implicit loci in GeoGebra to investigate Euler's inequality.
This well known statement can also be approached by a mixture of symbolic and numerical
observations. Our experiments are clearly not acceptable as a new way of proof, but
steps to claim promising conjectures. For other investigations of classic or new statements---that is,
to generalize geometric equations or inequalities---this kind of approach may be hopefully fruitful.

Also we highlight that a better approach might be to use real geometry and quantifier
elimination. To find the most efficient way to formalize and prove Euler's inequality and
present it in an adequate form in a dynamic geometry software tool is an on-going work of the authors.

As a final comment we demonstrated how Euler's inequality can be observed by using a new
tool, namely CindyJS. Here we had to write a few lines of program code.

\section{Acknowledgments}

The authors thank Tom\'as Recio for his helpful comments on the first version of this paper.

The first author was partially supported by a grant MTM2017-88796-P from the
Spanish MINECO (Ministerio de Economia y Competitividad) and the ERDF
(European Regional Development Fund).

\bibliography{kovzol,external}

\begin{thebibliography}{10}

\bibitem{ART-ISSAC2016}
Ab\'{a}nades, M., Botana, F., Kov\'{a}cs, Z., Recio, T., S\'{o}lyom-Gecse, C.:
\newblock Development of automatic reasoning tools in {G}eo{G}ebra.
\newblock ACM Commun. Comput. Algebra \textbf{50} (2016)  85--88

\bibitem{chou}
Chou, S.C.:
\newblock Mechanical geometry theorem proving.
\newblock Kluwer Academic Publishers Norwell, MA, USA (1987)

\bibitem{CollinsHong}
Collins, G.E., Hong, H.:
\newblock Partial Cylindrical Algebraic Decomposition for quantifier elimination.
\newblock Journal of Symbolic Computation \textbf{12}(3) (1991) 299--328

\bibitem{recio-dalzotto}
Dalzotto, G., Recio, T.:
\newblock On protocols for the automated discovery of theorems in elementary
  geometry.
\newblock Journal of Automated Reasoning \textbf{43} (2009)  203--236

\bibitem{cindyjs}
von Gagern, M.,
Kortenkamp, U.
Richter-Gebert, J.,
Strobel, M.:
\newblock {CindyJS}. Mathematical Visualization on Modern Devices.
\newblock In: G. M. Greuel, T. Koch T., P. Paule, A. Sommese (eds), Mathematical Software -- ICMS 2016.
Lecture Notes in Computer Science, vol 9725. Springer, Cham (2016)

\bibitem{ACA2015}
Ha\v{s}ek, R., Kov\'acs, Z., Zahradn\'{\i}k, J.:
\newblock Contemporary interpretation of a historical locus problem with the
  use of computer algebra.
\newblock In: Kotsireas, I.S., Mart\'{\i}nez-Moro, E., eds.: Applications of
  Computer Algebra: Kalamata, Greece, July 20--23 2015. Volume 198 of Springer
  Proceedings in Mathematics \& Statistics.
\newblock Springer (2017)

\bibitem{gg}
Hohenwarter, M.:
\newblock {G}eo{G}ebra: Ein {S}oftwaresystem f\"ur dynamische {G}eometrie und
  {A}lgebra der {E}bene.
\newblock Master's thesis, Paris Lodron University, Salzburg, Austria (2002)

\bibitem{cinderella}
Kortenkamp, U.:
Foundations of dynamic geometry. Doctoral Thesis, ETH Z\"urich (1999)

\bibitem{GiacGG-RICAM2013}
Kov\'acs, Z., Parisse, B.:
\newblock Giac and {G}eo{G}ebra -- improved {G}r\"obner basis computations.
\newblock In Gutierrez, J., Schicho, J., Weimann, M., eds.: Computer Algebra
  and Polynomials. Lecture Notes in Computer Science.
\newblock Springer (2015)  126--138

\bibitem{gg-art-doc-gh}
Kov\'acs, Z., Recio, T., V\'elez, M.P.:
\newblock gg-art-doc ({G}eo{G}ebra {A}utomated {R}easoning {T}ools. {A}
  tutorial).
\newblock A GitHub project (2017) \url{https://github.com/kovzol/gg-art-doc}.

\bibitem{mcs}
Kov\'acs, Z.:
\newblock Real-time animated dynamic geometry in the classrooms by using fast
  {G}r\"obner basis computations.
\newblock Mathematics in Computer Science \textbf{11} (2017)

\bibitem{ACA2017}
Kov\'acs, Z.:
\newblock Achievements and Challenges in Automatic Locus and Envelope Animations in Dynamic Geometry.
\newblock Mathematics in Computer Science (2018)
\url{https://doi.org/10.1007/s11786-018-0390-0}

\bibitem{extraverification}
Losada, R.:
\newblock El color din\'amico en {G}eo{G}ebra.
\newblock La Gaceta de la Real Sociedad Matem\'atica Espa\~nola \textbf{17}
  (2014)

\bibitem{SteinerLehmus}
Losada, R., Recio, T., Valcarce, J.L.:
\newblock On the automatic discovery of {S}teiner-{L}ehmus generalizations.
\newblock In: Proceedings of ADG'2010, Lecture Notes in Computer Science.
\newblock Springer, M\"unchen (2010)  171--174

\bibitem{cindyjs2}
Montag, A.,
Richter-Gebert, J.:
\newblock Bringing Together Dynamic Geometry Software and the Graphics Processing Unit.
\newblock arXiv:1808.04579 (2018)

\bibitem{Robu}
Robu, J.:
\newblock Automated Proof of Geometry Theorems Involving Order Relation in
the Frame of the Theorema Project.
\newblock In: Proceedings of the International
Conference on Knowledge Engineering, Principles and Techniques
(KEPT2007), Cluj-Napoca, Romania (2007) 307--315

\bibitem{acnode}
Wikipedia:
\newblock Acnode --- {W}ikipedia{,} the free encyclopedia (2016) [Online;
  accessed 10-July-2017].

\end{thebibliography}

\end{document}